\documentclass[twocolumn,resetfootnote]{aastex701}

\usepackage{graphicx}
\usepackage{subcaption}
\usepackage{placeins}
\usepackage{amsmath}

\begin{document}

\title{Pulsar glitches triggered by quakes when stress reaches a threshold}
%

\author{Zixuan Zhou}
\affiliation{School of Physics and State Key Laboratory of Nuclear Physics and Technology, Peking University, Beijing 100871, China}
\email{2401110291@stu.pku.edu.cn}

\author{Ruipeng Lu}
\affiliation{School of Earth and Space Sciences, Peking University, Beijing 100871, China}
\email{luruipeng@pku.edu.cn}

\author{Weiyang Wang}
\affiliation{School of Astronomy and Space Science, University of Chinese Academy of Sciences, Beijing 100049, China}
\email{wywang@ucas.ac.cn}

\author{Han Yue}
\affiliation{School of Earth and Space Sciences, Peking University, Beijing 100871, China}
\email{yue.han@pku.edu.cn}

\author{Renxin Xu}
\affiliation{School of Physics and State Key Laboratory of Nuclear Physics and Technology, Peking University, Beijing 100871, China}
\affiliation{Kavli Institute for Astronomy and Astrophysics, Peking University, Beijing 100871, China}
\email{r.x.xu@pku.edu.cn}

\correspondingauthor{Han Yue, Renxin Xu}

\begin{abstract}
%
The starquake model of pulsar glitches is revisited in light of accumulated observational data.
%
A statistical comparison between the time-predictable and the magnitude-predictable models, borrowed from seismology and grounded in the classical spring-slider framework, is performed against a sample of 245 glitch events from 9 pulsars.
%
%
It is found that the data supports the time-predictable model, i.e., a quake resulting in a glitch could be triggered when the developed stress reaches a critical threshold, $\tau_{\rm c}$, while the remaining elastic energy (correspondingly the residual stress, $\tau_{\rm r}$) after quakes is not fixed.
%
%
The critical stress could be the order of $\tau_{\rm c}\simeq (10^{34}-10^{35})\ {\rm dyn\,cm^{-2}}$ if the residual stress after a quake with maximum stress drops could be considered negligible (i.e., $\tau_{\rm r}\simeq 0$), to explain the glitch amplitudes and the intervals. This $\tau_{\rm c}$-value is the order of the shear modulus of strangeon stars, but would be several orders higher than that of conventional neutron stars.
%
%
\end{abstract}



\section{Introduction}\label{sec:intro}

In the era of \cite{1932PhyZS...1..285L}, it was thought that electrons, photons, neutrinos and nucleons (protons and neutrons\footnote{%
It is worth noting that there is a widespread misconception that L.~D. Landau proposed the concept of neutron stars following the discovery of the neutron. Please refer to the review by~\cite{2013PhyU...56..289Y} for more information on this.
}) %
were fundamentally point particles.
The basic units of both the atomic nucleus and Landau's gigantic nucleus are thus nucleons, but to be isospin-balanced in the former while neutron-rich in the latter due to the symmetry energy and the Fermi-Dirac statistics.
Consequently, superfluidity and vortex dynamics~\citep[][for a recent and comprehensive review]{2026arXiv260901022L} were soon proposed to explain the pulsar glitches discovered~\citep{1969Natur.222..228R}.
However, this is the era of the standard model of particle physics, and nucleons are certainly not point particles, but are composed of light-flavoured quarks (up, down and strange), despite the fact of zero strangeness there.
It is, therefore, very natural to revisit the glitch dynamics if the basic units of the gigantic nucleus differ from nucleons, and the concept of strange matter~\citep[e.g.,][]{xia_strange_2025} is one of the competing alternatives, and glitch models should then be re-investigated.
Here, in a simple spring-slider model, we examine pulsar glitches from the perspective of accumulated observational data, and find that glitches may be manifestations of starquakes, which are triggered when stress reaches a critical threshold, though the underlying mechanism is discussed speculatively.

As one of the four major astronomical discoveries of the 1960s, pulsars provide a testbed for some of the densest matter in the Universe, and their glitches, which are sudden increases in rotation frequency, can be used to study the interior physics of compact stars.
Frankly speaking, because of the challenging problem of the fundamental strong interaction at the low-energy regime, the nature of these stars remains controversial, to be either conventional neutron stars or strange stars~\citep{2005PrPNP..54..193W}.
Nevertheless, this topic is currently focused on neutron star models, and comprehensive reviews of glitch observations, theoretical models, and their physical interpretations can be found in the literature~\citep[e.g.,][]{haskell_models_2015,antonopoulou_pulsar_2022,zhou_pulsar_2022}.

Although the starquake model for conventional neutron stars (NSs) was also proposed at the very beginning~\citep{ruderman_neutron_1969, baym_neutron_1971}, it suffers from the well-known ``Vela glitch crisis'' and is then unable to reproduce the glitches as large and frequent as observed.
Nonetheless, this simple model could be revived if pulsars are in a globally solid state, i.e., the so-called solid strange stars~\citep{2004APh....22...73Z}.

This story dates back to an early conjecture concerning the state of absolutely stable strong matter at supra-nuclear densities, as proposed by~\citet{witten_cosmic_1984}.
In addition to quarks being the building blocks, quark clusters with strangeness (renamed ``strangeons'') could alternatively be the basic units in the ground state of strong matter, and pulsar-like compact stars would then be strangeon stars~\citep[SnSs,][]{xu_solid_2003,xiaoyu_strangeon_2017,qi_strangeon_2025} if the stronger coupling between quarks, which is non-perturbative in quantum chromodynamics, is taken seriously.
A key feature of this model is the global solidification of the star after cooling, making it a natural host for starquakes~\citep{2004APh....22...73Z,2014MNRAS.443.2705Z}. This framework has been progressively developed in a series of papers to confront various glitch observations. In Paper I~\citep{lai_pulsar_2018}, a comprehensive starquake picture was established for SnSs, demonstrating that the inner motion during a glitch can be decomposed into a non-recoverable plastic flow and a recoverable elastic motion, in order to explain the observed relation between the recovery coefficient $Q$ and glitch size. Paper II~\citep{wang_pulsar_2020} investigated the general glitch activity of normal radio pulsars within this model, providing constraints on the shear modulus of strangeon matter. In Paper III~\citep{lai_pulsar_2023}, the post-glitch recovery process was modeled as a viscous flow of matter toward the cracked equatorial plane, naturally reproducing the exponential recovery and uncovering a correlation between the cracking depth and the glitch size.
%
Recently, due to the stiff equation of state, a rapidly rotating SnS is found to have an ergosphere~\citep{Xia_2026}, resulting in extractable rotational energy on the order of $0.01 M_\odot$ before collapsing into a black hole, to be a potential candidate for the central engines of $\gamma$-ray bursts involving long-lived magnetars~\citep[e.g.,][]{1998A&A...333L..87D,2006ApJ...642..354Z}.

Starquakes occur naturally in SnSs and NSs, and a quantitative theoretical foundation for the quakes was established by~\citet{lu_quakes_2023} who performed three-dimensional elastic deformation calculations.
This work considered the stress-loading field from rotation deceleration and determined that the dominant mode of failure should be strike-slip faulting near the equator. A key result is that starquakes in the SnS model might be more plausible for explaining the observed glitch amplitudes compared to the NS model. In addition to the spin-down driven stress accumulation, magnetic field evolution may also drive starquakes, especially in magnetars, where Hall drift in the crust generates magnetic stresses that trigger avalanches of crustal failures \citep{li_magnetar_2016}.

The SnS quake model can be further combined with seismological predictability analyses to address the long-term patterns of stress accumulation and release that trigger the glitch sequence.
In seismology, the predictability of earthquake cycles is a fundamental probe of fault mechanics. The classical spring-slider model~\citep{burridge_model_1967} gives rise to three ideal predictable models: fully predictable model, time-predictable model, and magnitude-predictable model. Applying this analysis to pulsar glitches provides a direct link between observed timing and size statistics and the underlying physics.

In this paper, we perform a statistical comparison of these seismological predictable models against a large sample of pulsar glitch data. By computing the root-mean-square error (RMSE) for both the time-predictable and magnitude-predictable models, and systematically testing different glitch size thresholds, we aim to determine which model offers a superior statistical description of the data in \S2. We further estimate the key physical parameter, the critical stress $\tau_{\mathrm{c}}$, for the analyzed pulsars in \S3 and discuss the implications of our findings for the physics of starquakes in \S4. We summarize our results in \S5.

\section{Methods} 
\subsection{The Spring-slider Scenario}

\begin{figure}[t]
\centering
\includegraphics[width=\columnwidth]{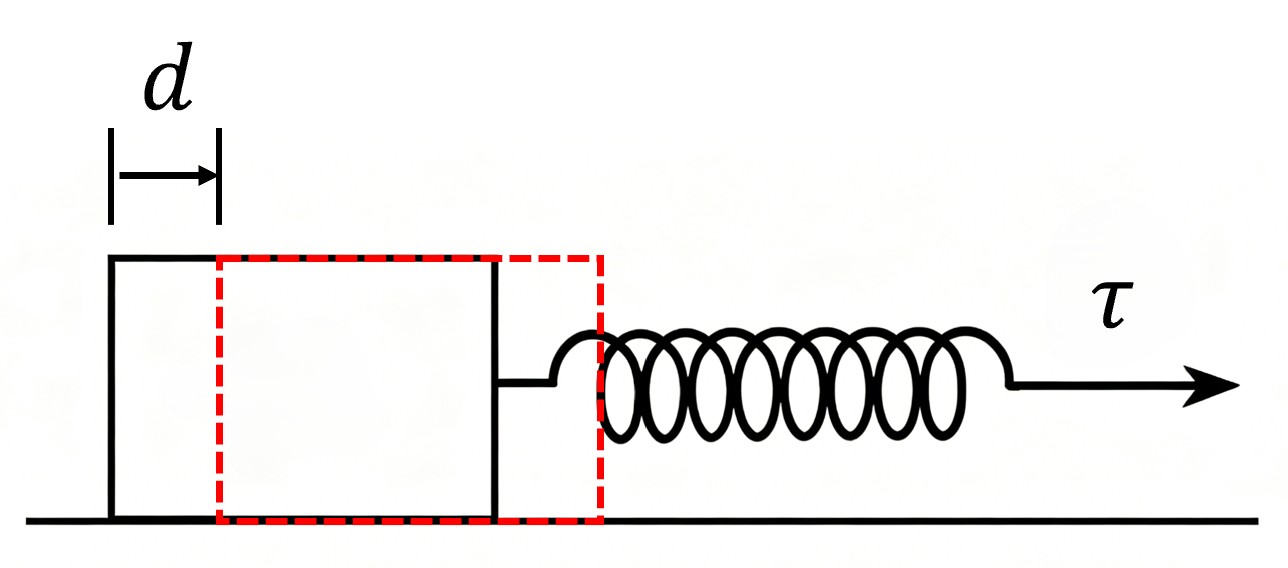}
\caption{A spring-slider model. A rectangular block, the slider, is connected to a spring. The system is driven by a constant shear stress, $\tau$, acting on the far end of the spring, which stretches the spring and then causes the slider to move by a displacement, $d$. The red dashed rectangle marks the displaced position of the slider after the spring stretches. As the spring elongates, elastic potential energy is accumulated until the static friction is overcome, triggering a sudden slip.
\label{fig:spring slider}}
\end{figure}

The spring-slider model is a classic mechanical analogue widely used to understand earthquake faulting \citep{burridge_model_1967}. Its basic components include an external force that pulls the spring at a constant velocity at the far end, providing a constant stress loading rate on the spring represented by the shear stress, $\tau$, a spring with a fixed elastic coefficient, and a slider with a certain mass and a bottom friction coefficient (Figure~\ref{fig:spring slider}). As the driving plate moves, the spring elongates by a displacement, $d$, accumulating elastic potential energy. When the spring force exceeds the maximum static friction force between the block and the surface, the block slips abruptly, releasing the stored energy as an earthquake.

We apply this model to understand pulsar glitches within the starquake scenario. Due to the conservation of rotational angular momentum of pulsars, one may have
\begin{equation}
\frac{{\rm d}\nu}{\nu} = \frac{{\rm d}\omega}{\omega} = -\frac{{\rm d}I}{I} .
\label{eq:conservation}
\end{equation}
A gravity-induced starquake tends to change a star's state so that it has lower gravitational energy, resulting in a decrease in the pulsar's moment of inertia,\footnote{%
An anti-glitch, i.e., increasing the moment of inertia, could also be possible if the magnetic stress in a compact object dominates the elastic deformation~\citep{2026arXiv260712285L}.
} $I$. %
Therefore, we adapt this model by equating the block displacement ($d$) to the change of the pulsar's moment of inertia (${\rm d} I$), leading then to the change in the rotational frequency, ${\rm d}\nu$, which could be detected by the telescope.

\subsection{Three Predictable Models}

\begin{figure}[t]
\centering
\includegraphics[width=\columnwidth]{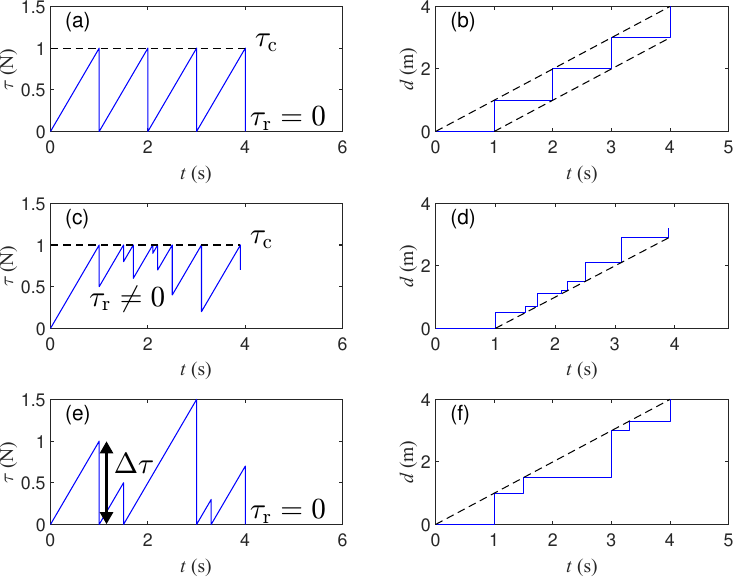}
\caption{Temporal evolution of accumulated stress $\tau$ (left column) and cumulative displacement $d$ (right column) for three predictable models in the spring-slider framework. Top row: fully predictable model, in which the critical stress $\tau_{\mathrm{c}}$ is fixed and the residual stress $\tau_{\mathrm{r}}$ after each quake is zero. Middle row: time-predictable model, with a constant critical stress $\tau_{\mathrm{c}}$ and a changing residual stress $\tau_{\mathrm{r}}$ after each event. Bottom row: magnitude-predictable model, with a changing critical stress $\tau_{\mathrm{c}}$ and a zero residual stress $\tau_{\mathrm{r}}$ after each quake. In the left panels, the dashed horizontal line marks the critical stress $\tau_{\mathrm{c}}$. In the right panels, the dashed diagonal lines indicate the expected linear relations for reference. \label{fig:three models}}
\end{figure}

The predictability of earthquake cycles within the spring-slider framework has been discussed\citep{carlson_dynamics_1994}.
In the framework of the basic spring-slider, the rupture interval, corresponding to earthquake period, of the slider depends on two factors: the rupture threshold and the stress drop per earthquake over time. The difference of these two factors leads to three kinds of predictable models. To better understand these models, temporal evolution of accumulated stress, $\tau$, and temporal evolution of cumulative displacement, $d$, for three models are shown in Figure~\ref{fig:three models}. 

\begin{enumerate}
\item {\em The fully predictable model}:
If both the rupture threshold, $\tau_{\rm c}$, and the stress drop (i.e., $\tau_{\rm c}-\tau_{\rm r}$, with $\tau_{\rm r}$ the residual stress after rupture) remain constant, the same stress accumulates during each interseismic period, and is released during each earthquake. Consequently, both the earthquake intervals and magnitudes are completely predictable.

\item {\em The time-predictable model}: The rupture threshold $\tau_{\rm c}$ remains constant, but the stress drop varies.
This implies that quakes occur when the accumulated stress reaches a critical threshold, while the residual elastic energy after a quake is nonzero and fluctuates from event to event.
The time of the next event is thus predictable, but the magnitude is not.
We will show that the data from pulsar glitches support this model.

\item {\em The magnitude-predictable model}: The residual stress after rupture is constant (e.g., $\tau_{\rm r}=0$), while the rupture threshold $\tau_{\rm c}$ varies.
The magnitude of the next event is thus predictable, but the time interval is not.

\end{enumerate}

\label{sec:methods}

\subsection{Origin Data Analysis}

\begin{figure*}[ht]
\centering
\includegraphics[width=0.9\textwidth]{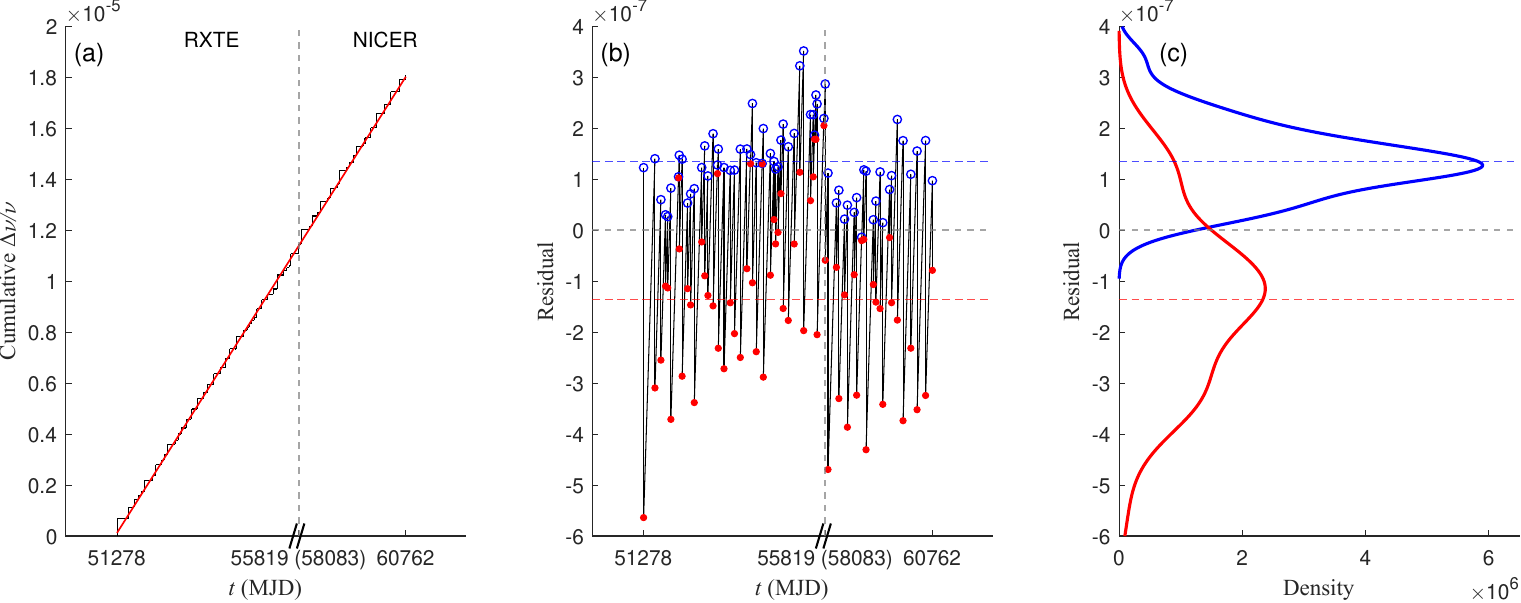}
\caption{Analysis of PSR J0537-6910 glitch data. (a) Cumulative relative frequency change $C(t)=\sum_{t_i\le t}(\Delta\nu/\nu)_i$ as a function of time, corresponding to the right  column of Figure~\ref{fig:three models}, and the red line is a least-squares linear fit. (b) Detrended residuals $\delta C(t)=C(t)-C_{\rm fit}(t)$ plotted against time separated by glitch phase, corresponding to the left  column of Figure~\ref{fig:three models}. Blue points are points before each glitch, which cluster tightly; red points are points after each glitch, showing a larger scatter. The smaller dispersion of the pre-glitch residuals supports the time-predictable model, while the larger spread of the post-glitch residuals disfavors the magnitude-predictable model. (c) Probability density functions of the two populations; the pre-glitch distribution is noticeably narrower. The asymmetry in residual dispersion indicates that the glitch sequence of this pulsar is more consistent with a time-predictable than with a magnitude-predictable behavior. \label{fig:residual density}}
\end{figure*}

PSR J0537-6910  is the fastest spinning young pulsar in the supernova remnant N157B in the Large Magellanic Cloud~\citep{wang_rosat_1998}. There are 68 glitches detected over the more than 20 yr of combined observations (1999–2011 with RXTE and 2017–2025 with NICER)~\citep{ middleditch_predicting_2006, antonopoulou_pulsar_2018, ferdman_glitches_2018, ho_return_2020, abbott_diving_2021, ho_timing_2022, ho_new_2026}.
To develop an intuition, we tested the difference between the time- and magnitude-predictable models using the original data of PSR J0537-6910, as this pulsar has glitched more frequently than any other up to now.

For each glitch event, we have the relative rotation frequency change $\Delta\nu/\nu$. The cumulative relative rotation frequency change is calculated by $C(t)=\sum_{t_i \le t} \Delta\nu_i/\nu$. Subplot~(a) of Figure~\ref{fig:residual density} displays cumulative relative rotation frequency change as a function of time. The red line presents a least-squares linear fit to all points. Then we examine the detrended residuals $\delta C(t) \equiv C(t) - C_{\rm fit}(t)$, separated according to whether the point is just before a glitch (blue points) or just after a glitch (red points). Subplot~(b) of Figure~\ref{fig:residual density} shows the time series of these residuals. The blue points cluster tightly with a mean of $+1.35\times10^{-7}$ and a standard deviation of $7.41\times10^{-8}$. In contrast, the red points exhibit a larger scatter: mean $= -1.35\times10^{-7}$, standard deviation $= 1.70\times10^{-7}$. The probability density functions of the two populations are compared in subplot~(c). The distribution of the blue points is apparently narrower than that of the red points.

This asymmetry in the dispersion of the detrended residuals indicates that the glitch of PSR J0537-6910 is more consistent with a time-predictable model than with a magnitude-predictable model.

\subsection{Data Source and Sample}

\begin{figure}[t]
\centering
\includegraphics[width=\columnwidth]{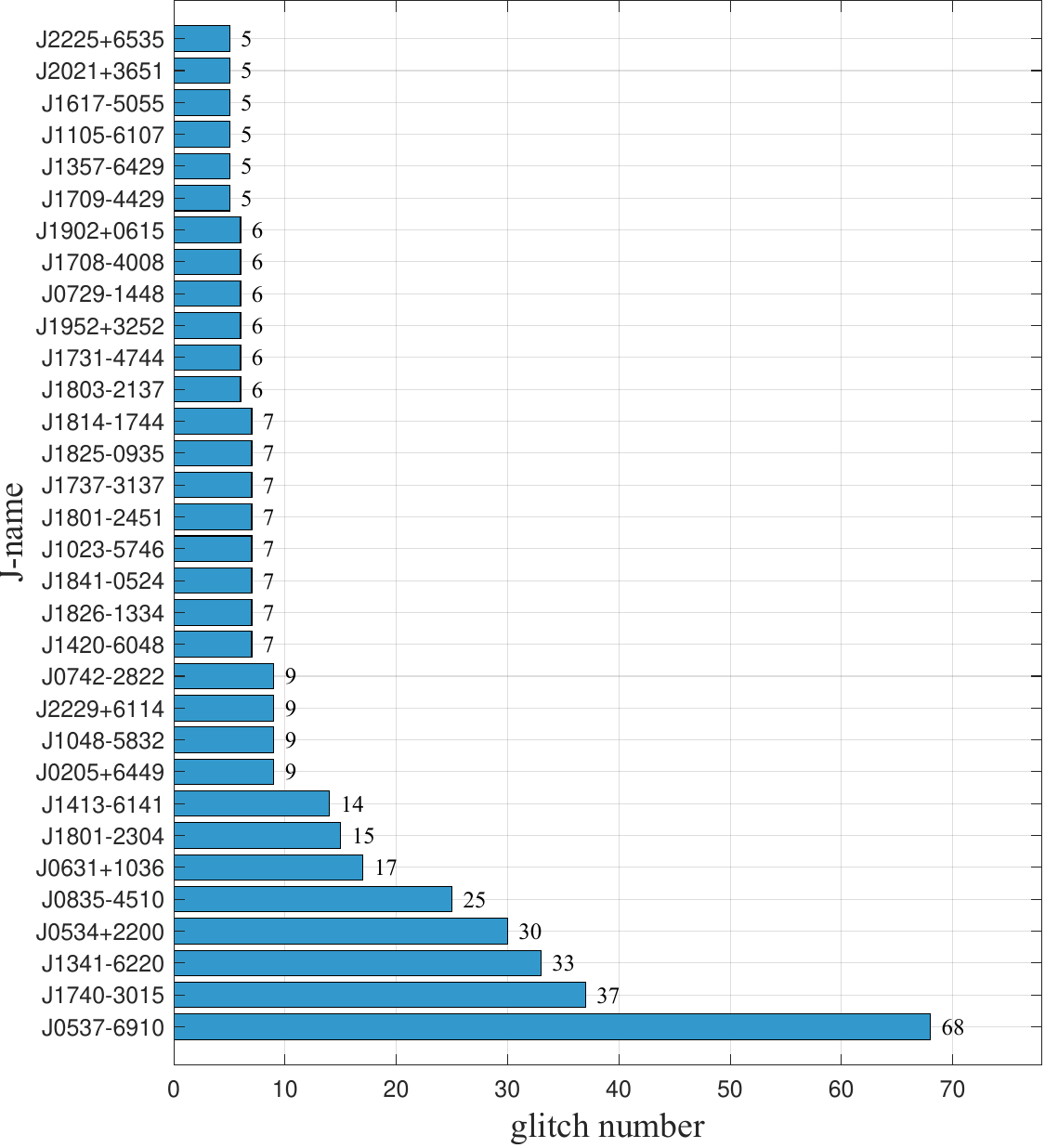}
\caption{Number of glitches per pulsar. Only pulsars with at least 5 recorded glitches are included. In this work, the nine pulsars with more than 10 recorded glitches are selected. \label{fig:glitch number}}
\end{figure}

The glitch data used in this analysis are retrieved from the ATNF Pulsar Catalogue Glitch Table. The database, accessible at \url{https://www.atnf.csiro.au/people/pulsar/psrcat/glitchTbl.html}, compiles published glitch parameters for pulsars. For this study, we extracted all entries containing a valid measurement for the fractional glitch size, $\Delta \nu / \nu$. Our final sample comprises 245 glitch events from 9 pulsars, as shown in Figure~\ref{fig:glitch number}.


\subsection{Calculating Root-Mean-Square Error}
\label{subsec:thresholding}

In observational glitch sequences,  a large glitch is often accompanied by numerous smaller events, corresponding to the main shock, foreshocks and aftershocks in earthquake. To test the predictable models, it is necessary to isolate these main events from the smaller accompanying ones. The spring-slider model describes the large-scale release of accumulated crustal stress. Very small glitches may represent localized, partial fractures or different physical processes. Therefore, our analysis focuses on identifying the large-scale starquakes that dominate the long-term spin evolution of the pulsar. We apply a relative threshold to each pulsar's glitch, filtering out those small events.

From the full ATNF glitch catalog, we identify all pulsars with more than 10 recorded glitches, which yields 9 candidate pulsars for this analysis. For each pulsar, we define a relative threshold based on its own historical glitch activity
\begin{equation}
    (\Delta \nu / \nu)_{\text{th}} = f \cdot \max\left[\left(\Delta \nu / \nu\right)_i\right] ,
    \label{eq:threshold}
\end{equation}
where \( \max\left[(\Delta \nu / \nu)_i\right] \) is the largest recorded glitch size for that pulsar, and \( f \) is the threshold fraction. We systematically test three distinct thresholds: \( f \) = 0.01, 0.05, 0.1. Any glitch with a size \( (\Delta \nu / \nu)_i < (\Delta \nu / \nu)_{\text{th}} \) is removed from the sequence. The remaining glitches constitute the principal glitch sequence. The time \( t_i \) and sizes \( (\Delta \nu / \nu)_i \) of these principal events are then used to calculate the glitch intervals \( \Delta t \) and size changes \( \Delta\nu \) for direct input into the time- and magnitude-predictable model.

For each pulsar, after discarding glitches with sizes below the relative threshold $f$, we obtain a sequence of $N_{\mathrm{g}}$ principal events characterized by their time $t_i$ and sizes $(\Delta\nu/\nu)_i$. The waiting time between consecutive glitches is defined as $\Delta t_i = t_{i+1} - t_i$. To evaluate two predictive models, we perform least-squares regressions on two different pairings of the data.

In the time-predictable model framework, the time of next glitch happening is expected to be proportional to the size of the previous event. Accordingly, we pair each waiting time $\Delta t_i$ with the size of the preceding glitch $\Delta\nu$, and fit a linear relation constrained to pass through the origin. The RMSE of this fit, denoted $\mathrm{RMSE}_{\mathrm{t}}$, quantifies the scatter around the predicted linear trend; a perfectly time-predictable sequence would yield $\mathrm{RMSE}_{\mathrm{t}}=0$.

Conversely, for the magnitude-predictable model,  the magnitude of next glitch is expected to be proportional to the time interval since the previous event. We therefore pair each waiting time $\Delta t_i$ with the size of the succeeding glitch, $\Delta\nu$, and perform a similar zero-intercept linear fit to obtain $\mathrm{RMSE}_{\mathrm{m}}$. In this case, a perfectly magnitude-predictable sequence would yield $\mathrm{RMSE}_{\mathrm{m}}=0$.

\section{Results} \label{sec:Results}

\begin{figure*}[t!]
\centering
\includegraphics[width=0.8\textwidth]{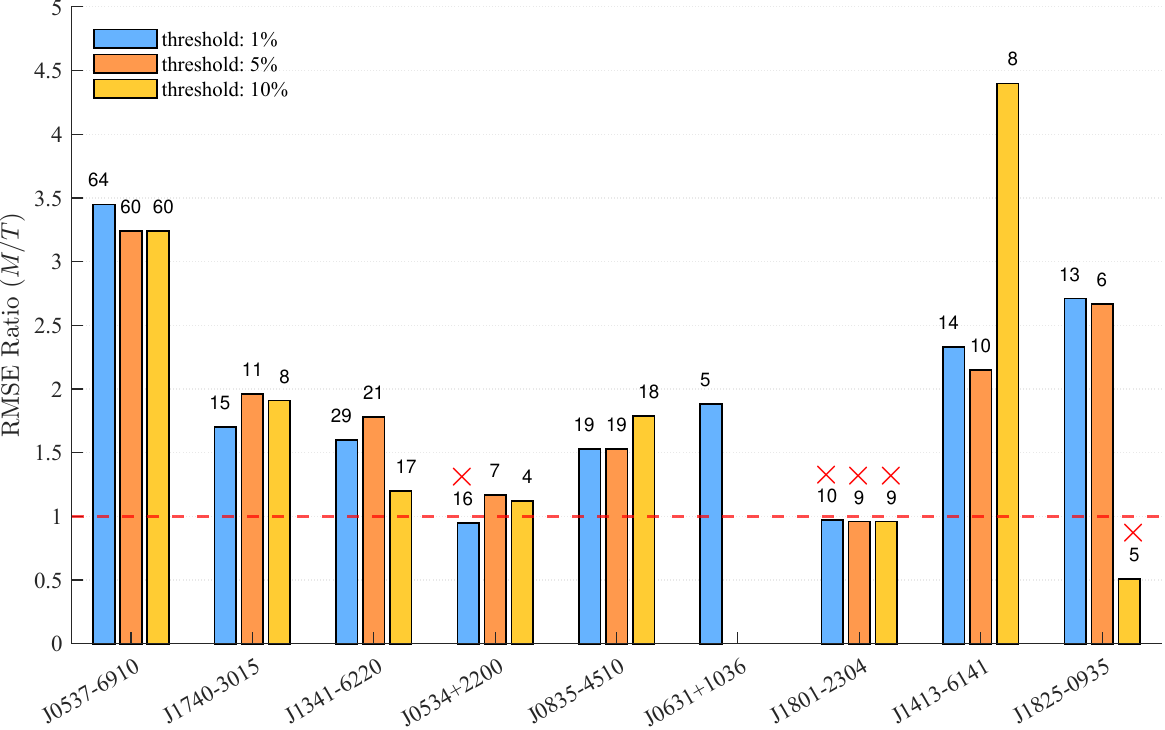}
\caption{Comparison of the time-predictable and magnitude-predictable models for 9 pulsars with at least 10 recorded glitches. The bars show the ratio ${\rm RMSE_m}/{\rm RMSE_t}$ at three relative thresholds: $f=1\%$, $5\%$, $10\%$ of the maximum glitch size. Numbers above each bar indicate the number of glitches $N_g$ retained after thresholding. The red dashed line marks unity; values above this line indicate that the time-predictable model yields smaller prediction errors. \label{fig:three thresholds results}}
\end{figure*}

\setlength{\tabcolsep}{8pt}
\begin{deluxetable*}{l *{13}{c}}[t!]
\tabletypesize{\scriptsize}
\tablewidth{\linewidth}
\tablecaption{Comparative analysis of glitch predictable models at 1\%, 5\%, and 10\% thresholds \label{tab:rmse_comparison}}
\tablehead{
\colhead{Pulsar} &
\colhead{$N_{\rm tot}$} &
\multicolumn{4}{c}{1\%} & \multicolumn{4}{c}{5\%} & \multicolumn{4}{c}{10\%} \\
\cline{3-6} \cline{7-10} \cline{11-14}
& &
\colhead{$N_{\rm g}$} & \colhead{$M$} & \colhead{$T$} & \colhead{$M/T$} &
\colhead{$N_{\rm g}$} & \colhead{$M$} & \colhead{$T$} & \colhead{$M/T$} &
\colhead{$N_{\rm g}$} & \colhead{$M$} & \colhead{$T$} & \colhead{$M/T$}
}
\startdata
J0537$-$6910 & 68 & 64 & 0.64 & 0.18 & 3.45 & 60 & 0.58 & 0.18 & 3.24 & 60 & 0.58 & 0.18 & 3.24 \\
J1740$-$3015 & 37 & 15 & 0.89 & 0.52 & 1.70 & 11 & 0.74 & 0.38 & 1.96 & 8 & 0.63 & 0.33 & 1.91 \\
J1341$-$6220 & 33 & 29 & 0.86 & 0.54 & 1.60 & 21 & 0.75 & 0.42 & 1.78 & 17 & 0.57 & 0.47 & 1.20 \\
J0534$+$2200 & 28 & 16 & 0.87 & 0.92 & 0.95 & 7 & 0.85 & 0.73 & 1.17 & 4 & 0.72 & 0.65 & 1.12 \\
J0835$-$4510 & 21 & 19 & 0.42 & 0.28 & 1.53 & 19 & 0.42 & 0.28 & 1.53 & 18 & 0.42 & 0.24 & 1.79 \\
J0631$+$1036 & 17 & 5 & 0.97 & 0.52 & 1.88 & 2 & —— & —— & —— & 2 & —— & —— & —— \\
J1801$-$2304 & 13 & 10 & 0.50 & 0.52 & 0.97 & 9 & 0.50 & 0.52 & 0.96 & 9 & 0.50 & 0.52 & 0.96 \\
J1413$-$6141 & 14 & 14 & 0.81 & 0.35 & 2.33 & 10 & 0.70 & 0.32 & 2.15 & 8 & 0.79 & 0.18 & 4.40 \\
J1825$-$0935 & 14 & 13 & 0.97 & 0.36 & 2.71 & 6 & 0.85 & 0.32 & 2.67 & 5 & 0.32 & 0.64 & 0.51 \\
\enddata
\tablecomments{$N_{\rm tot}$: Total glitches (ATNF). $N_{\rm g}$: Glitch number satisfying threshold. $M$: RMSE$_{\rm m}$, $T$: RMSE$_{\rm t}$, $M/T$: RMSE$_{\rm m}$/ RMSE$_{\rm t}$. }
\end{deluxetable*}

For pulsars exhibiting more than 10 recorded glitches, we performed a quantitative comparison between the time-predictable and magnitude-predictable models. For each glitch sequence, the RMSE was computed for both the time-predictable model (RMSE$_{\rm t}$) and the magnitude-predictable model (RMSE$_{\rm m}$). The ratio $\mathrm{RMSE}_{\mathrm{m}}/\mathrm{RMSE}_{\mathrm{t}}$ provides a direct metric for comparing the two models. A ratio greater than unity indicates that the time-predictable model produces smaller residuals and thus offers a more accurate description of the observed glitch sequence. On the other hand, a ratio less than unity indicates that the magnitude-predictable model fits better. This calculation was repeated across three threshold levels to assess the model performance under different criteria for what constitutes a significant glitch. The comparative results are visualized in Figure~\ref{fig:three thresholds results}  and summarized in Table~\ref{tab:rmse_comparison}.

Figure~\ref{fig:three thresholds results} displays the ratio RMSE$_{\rm m}$/RMSE$_{\rm t}$ for each pulsar at three thresholds. The numbers above each bar show the number of glitches ($N_g$) used in the analysis for that pulsar and threshold. Values greater than unity, indicated by the red dashed line, signify that the time-predictable model yields a smaller prediction error. Across all three thresholds, RMSE$_{\rm t}$ is found to be systematically smaller than RMSE$_{\rm m}$ for the vast majority of pulsars in our sample. Out of the 25 comparisons, the time-predictable model yields a lower error in 20 cases, corresponding to 80\% of the total. This statistically dominant trend indicates that  for this population of frequently glitching pulsars the time-predictable model provides a consistently better description of the glitch occurrence pattern than the magnitude-predictable model.

The superior performance of the time-predictable model aligns naturally with the starquake picture within the strangeon star model. In this framework, glitches are interpreted as discrete energy-release events resulting from the fracturing of the rigid strangeon crust. The process is primarily driven by the steady spin-down of the star, which progressively increases the stress within the crust until a critical threshold is reached, triggering a quake. This stress-accumulation mechanism inherently predicts a quasi-periodic pattern in glitch timing, which is precisely what the time-predictable model captures.

While these results strongly favor the time-predictable model, we note a small subset of pulsars shows the opposite behavior. This minority is confined to specific thresholds in J0534$+$2200 and J1825$-$0935. This may indicate the presence of a different glitch mechanism in some objects, the influence of unresolved small glitches, or more complex dynamics process that deviates from a simple time-predictable model. Future work with expanded glitch samples and multi-messenger constraints will be crucial to explore these exceptions in detail.

\begin{figure}[t!]
\centering
\includegraphics[width=\columnwidth]{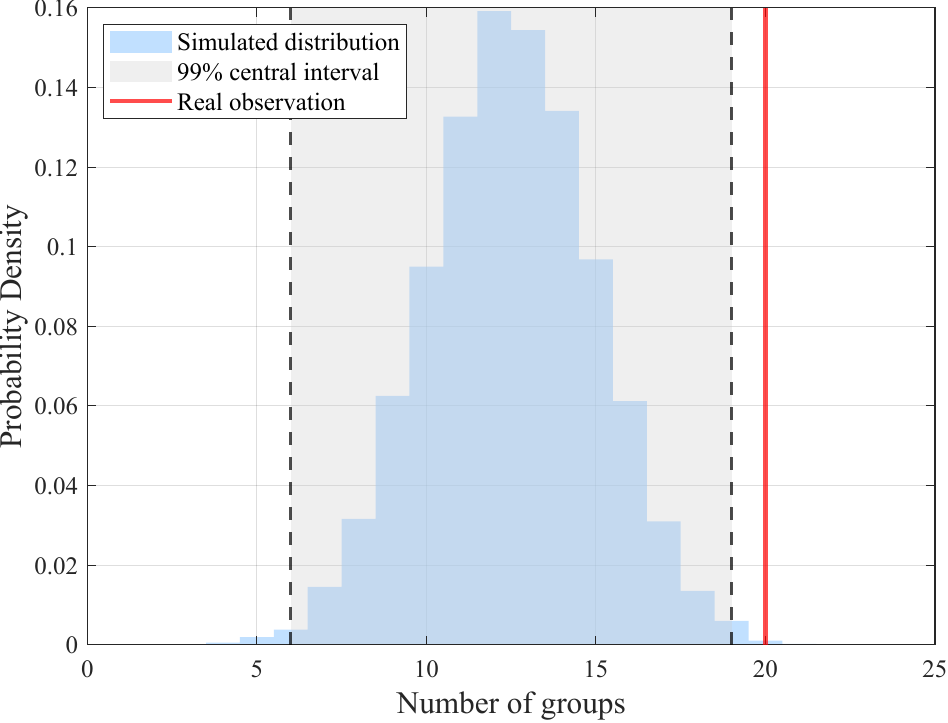}
\caption{Distribution of the number of groups in which the time-predictable model outperforms the magnitude-predictable model, obtained from 10,000 realizations of completely random data. The gray shaded region marks the 99\% central interval of the distribution; the red vertical line indicates the real observed value of 20 groups. The resulting distribution has a mean of 12.50 and a standard deviation of 2.48. Only 12 out of 10,000 reached or exceeded the observed value of 20.}
\label{fig:random simulate}
\end{figure}

To assess whether the observed superiority of the time-predictable model could be a statistical fluke, we performed two sets of control experiments using purely random data, in which both waiting times and glitch sizes were independently drawn from a uniform distribution \((0,1) \). Firstly, we generated 10000 glitches and computed the RMSE ratio \( R = \mathrm{RMSE}_{\mathrm{m}} / \mathrm{RMSE}_{\mathrm{t}} \) using the same normalized RMSE defined in Section~2.3. The resulting ratio is \( R = 1.000849 \), with \( \mathrm{RMSE}_{\mathrm{t}} = 0.6574 \) and \( \mathrm{RMSE}_{\mathrm{m}} = 0.6579 \). This value is virtually unity, indicating that for purely random data the two models are perfectly symmetric in their predictive power, and no significant asymmetry can arise from stochastic fluctuations alone. Secondly, we preserved the sample size \( N_g \) of each of the 25 groups in Table~\ref{tab:rmse_comparison}, and for each group we generated completely random waiting times and glitch sizes independently. For each realization, we counted how many of the 25 groups yielded \( R > 1 \). We repeated this procedure 10,000 times, and the resulting distribution is shown in Figure~\ref{fig:random simulate}. The distribution has a mean of \( 12.50 \) and a standard deviation of \( 2.48 \). The gray shaded region in the figure marks the 99\% central interval of the distribution, bounded by the 0.5\% and 99.5\% quantiles, which covers the range where 99\% of random realizations fall.  In the real data, the time-predictable model is better in 20 out of 25 groups, indicated by the red vertical line, and its value lies well outside this interval. In the 10,000 random realizations, only 12 cases reached or exceeded this value, showing the probability of obtaining 20 or more successes in 25 groups is only \( 0.12\% \). In conclusion, these two control experiments demonstrate that the 20/25 superiority of the time-predictable model cannot be attributed to random fluctuations, and the behavior revealed in Figure~\ref{fig:three thresholds results} is statistically robust and reflects genuine physics, which we turn to discuss in the next section.

\section{Discussion} \label{sec:discussion}

The statistical results presented in Section~3 demonstrate that the time-predictable model provides a superior description of the observed glitch sequences for the majority of pulsars in our sample. This suggests that the glitch recurrence is primarily governed by a failure threshold. We utilize the elastic deformation theory developed for strangeon stars by \citet{lu_quakes_2023} to estimate the stress drop associated with individual glitch events from their observed amplitudes.

In the framework of \citet{lu_quakes_2023}, the fractional change in rotation frequency during a starquake is related to the strain drop \(\Delta\epsilon\) by their Equation~(15)
\begin{equation}
\frac{\Delta\nu}{\nu} = \frac{\chi \Delta\epsilon \, \mu R^3 \Gamma}{I},
\label{eq:lu15}
\end{equation}
where \(\mu\) is the shear modulus, \(R\) is the stellar radius, \(I\) is the moment of inertia, \(\Gamma\) is the efficiency factor linking the seismic moment to the change in moment of inertia, and \(\chi = (r/R)^3\) is the dimensionless rupture volume ratio, with \(r\) being the characteristic rupture size.

\begin{figure*}[htbp]
\centering
\includegraphics[width=0.7\textwidth]{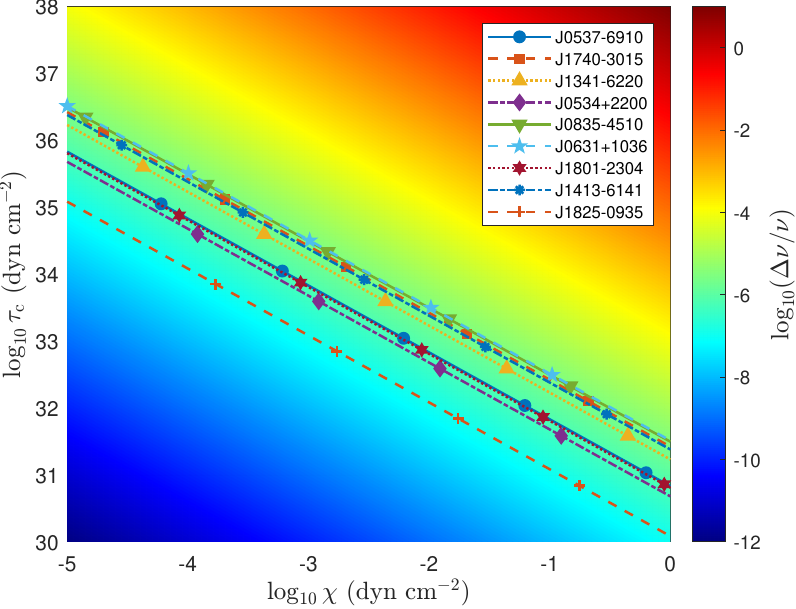}
\caption{Relation between the critical stress \(\tau_{\rm c}\) and the rupture volume ratio $\chi$ for the nine pulsars analyzed in this work. The colour background encodes the glitch size $\Delta\nu/\nu$ on a logarithmic scale. The lines trace Equation~(\ref{eq:stressdrop_main}) for each pulsar. Different colors, line styles, and marker symbols are used to distinguish individual pulsars, as listed in the legend.}
\label{fig:chi_tau}
\end{figure*}

For an elastic medium, the stress drop \(\Delta\tau\) is related to the strain drop by Hooke's law
\begin{equation}
\Delta\tau = \mu \, \Delta\epsilon .
\label{eq:hooke}
\end{equation}
We can obtain the relation between the observed glitch amplitude and the released stress drop
\begin{equation}
\Delta\tau = \frac{(\Delta\nu/\nu) \, I}{\chi R^3 \Gamma}
= 10^{34} (\Delta\nu/\nu)_6 \, I_{45}\, \chi_3^{-1} R_6^{-3} \Gamma_{10}^{-1}
\ {\rm dyn\,cm^{-2}},
\label{eq:stressdrop_main}
\end{equation}
where
$(\Delta\nu/\nu)= (\Delta\nu/\nu)_6\times 10^{-6}$, 
$R=R_6\times 10^6\ {\rm cm}$, 
$I = I_{45}\times 10^{45}\ {\rm g\,cm^2}$,
$\chi = \chi_3 \times 10^{-3}$, 
$\Gamma = \Gamma_{10} \times 10^{-10}\ {\rm s^2}$.

For the neutron star cases, we can similarly get the released stress drop
\begin{equation}
\Delta\tau = 10^{35} (\Delta\nu/\nu)_6 \, I_{45}\, \chi_3^{-1} R_6^{-3} \Gamma_{9}^{-1}
\ {\rm dyn\,cm^{-2}},
\label{eq:stressdrop_ns}
\end{equation}
where
$\Gamma = \Gamma_{9} \times 10^{-9}\ {\rm s^2}$, other parameters are the same.

We select the pulsars listed in Table~\ref{tab:rmse_comparison}. For each selected pulsar, we identify the maximum observed glitch amplitude, denoted as \((\Delta\nu/\nu)_{\max}\) under the assumption that the residual stress after a quake with maximum stress drop could be considered negligible (i.e., $\tau_{\rm r}\simeq 0$). Therefore, the derived stress drop can be taken as a proxy for the critical stress $\tau_{\rm c}$ required to trigger a starquake in the strangeon star. Figure~\ref{fig:chi_tau} displays the relationship between $\tau_{\rm c}$  and $\chi$ for all nine pulsars. The color background indicates the corresponding glitch size on a logarithmic scale. The solid lines trace Equation~(\ref{eq:stressdrop_main}) for each pulsar. Different colors, line styles, and marker symbols distinguish individual pulsars as listed in the legend.

The strain drops for starquakes lie in the range of $10^{-5}$ to $10^{-1}$ \citep{horowitz_breaking_2009,baiko_breaking_2018,ruderman_neutron_1991}.
Combined with Equation~\eqref{eq:hooke}, the required stress drop for glitch size similar to Vela is generally within the range of the theoretical shear modulus of strangeon matter, $\mu \sim 10^{32}$ to $10^{35}\ {\rm dyn\,cm^{-2}}$ (\citealt{xu_solid_2003}; \citealt{lu_quakes_2023}). \footnote{%
We note that the vertical axis label in the right-hand panel of Figure~6 of \citet{lu_quakes_2023} should be $\ln$ rather than $\lg$.%
}
This consistency suggests that the strain energy stored in a globally solidified strangeon star is sufficient to power the observed large glitches.
In contrast, for conventional neutron stars the shear modulus required to produce large glitches would be several orders of magnitude larger than the neutron star crustal shear modulus of $10^{30}\ {\rm dyn\,cm^{-2}}$ estimated by \citet{baym_neutron_1971}. This discrepancy may suggest that the elastic energy stored in the crust of a conventional neutron star is insufficient to explain the large glitches observed, a fact that was first noted in the 1970s.

\section{Conclusions} \label{sec:conclusion}

In this work, we performed the first comparison of seismological predictable models using a sample of 245 glitch events from 9 pulsars. By computing the RMSE for both models and testing different glitch size thresholds, we identified the model that offers a superior statistical description of the glitch occurrence pattern. Furthermore, we utilized the elastic deformation theory for strangeon stars to estimate the stress drop associated with the largest glitch of each pulsar, under the assumption of complete stress release. Our main conclusions are as follows:

\begin{enumerate}
\item The time-predictable model provides a consistently better description of glitch sequences than the magnitude-predictable model. Across three threshold levels, $1\%$, $5\%$, and $10\%$ of the maximum glitch size, the time-predictable model yields a smaller RMSE in 20 out of 25 comparisons. This statistically dominant trend suggests that glitch recurrence in frequently glitching pulsars is primarily governed by a critical stress threshold.

\item The dimensionless form of the stress drop relation, $\Delta\tau = 10^{34} (\Delta\nu/\nu)_6\, I_{45}\, \chi_3^{-1} R_6^{-3} \Gamma_{10}^{-1}\ {\rm dyn\,cm^{-2}}$, provides a convenient tool for estimating the critical stress from observed glitch sizes, without relying on the uncertain shear modulus of strangeon matter.

\item We estimated the critical stress for nine pulsars. For a starquake with $\chi \sim 10^{-3}$, the required stress drop to produce a Vela-sized glitch  is $\Delta\tau \sim 10^{34}\ {\rm dyn\,cm^{-2}}$, which falls within the theoretical shear modulus range of strangeon matter, $\mu \sim 10^{32}$ to $10^{35}\ {\rm dyn\,cm^{-2}}$. This consistency implies that the strain energy stored in a globally solidified strangeon star is sufficient to power the observed large glitches.

\end{enumerate}

Future work with expanded glitch samples, longer observational baselines, and improved theoretical modeling of the elastic properties of strangeon matter will be essential to further constrain the parameters of the starquake model and to distinguish between competing scenarios of pulsar glitch mechanisms.

\begin{acknowledgments}
This work was supported by the National SKA Program of China (2020SKA0120100, 2020SKA0120300) and  the National Natural Science Foundation of China (Grant No. 42174059).
We are grateful to Dr. Xinyu Li from Tsinghua University for his helpful discussion about $B$-field driven starquakes in compact objects.
\end{acknowledgments}


\end{document}